\documentclass[preprint,12pt]{elsarticle}

\usepackage{amssymb}

\journal{Annals of Physics}

\begin{document}

\begin{frontmatter}



\title{Propagation of light and retarded time of radiation in a strong gravitational wave\tnoteref{t1}}
\tnotetext[t1]{The study was supported by the Russian Science Foundation, grant \mbox{No.\,23-22-00343},
{https://rscf.ru/en/project/23-22-00343/}}


\author[1,2,3]{Konstantin E. Osetrin\corref{cor1}}%
\ead{osetrin@tspu.edu.ru}

\author[1]{Vladimir Y. Epp
}
\ead{epp@tspu.edu.ru}

\author[4,5]{Sergey V. Chervon
}
\ead{chervon.sergey@gmail.com}

%

\cortext[cor1]{Corresponding author}

\affiliation[1]{organization={Faculty of Physics and Mathematics, Tomsk State Pedagogical University},
            addressline={Kievskaya street 60}, 
            city={Tomsk},
            postcode={634061}, 
            country={Russia}}

\affiliation[2]{organization={International Laboratory for Theoretical Cosmology, Tomsk State University of Control Systems and Radioelectronics},
            addressline={Lenina Prospect 40}, 
            city={Tomsk},
            postcode={634050}, 
            country={Russia}}
            
\affiliation[3]{organization={Faculty of Physics, Tomsk State University},
            addressline={Lenina Prospect 36}, 
            city={Tomsk},
            postcode={634050}, 
            country={Russia}}
            
\affiliation[4]{organization={Laboratory of gravitation, cosmology and astrophysics, Ulyanovsk State Pedagogical
University},
            addressline={Lenin’s Square 4/5}, 
            city={Ulyanovsk},
            postcode={432071}, 
            country={Russia}}

\affiliation[5]{organization={Physics Department, Bauman Moscow State Technical University},
            addressline={2-nd Baumanskaya street 5}, 
            city={Moscow},
            postcode={105005}, 
            country={Russia}}

\begin{abstract}
For the strong gravitational wave model, an explicit transformation is obtained from a privileged coordinate system with a wave variable to a synchronous reference frame with separation of time and space variables. In a synchronous reference frame, a general form of the gravitational wave metric, solutions to the equations of trajectories for test particles in the Hamilton-Jacobi formalism, a solution to the eikonal equation for radiation, and a form of equations for the light ''cone'' of an observer in a gravitational wave were found. Using the obtained relations, the form ''the retarded time of radiation'' in the gravitational wave was found. The general relations obtained can be applied both in Einstein's general theory of relativity and in modified theories of gravity. The obtained relations were applied in the work for an exact model of a gravitational wave in the Bianchi type VI universe based on an exact solution of Einstein’s vacuum equations.
\end{abstract}




\end{frontmatter}



\section{Introduction}
The study of the trajectories of light rays in a gravitational field is a classical problem in the theory of gravitation, since it forms the mathematical basis for observational astronomy \cite{LandauEng1}. In practice, due to their complexity, such problems are mainly solved on the basis of perturbative and 
numerical methods~\cite{MUKHANOV1992203,Ma19957}. Therefore, obtaining exact mathematical models of the propagation of radiation in a gravitational field is a significant scientific result, and such exact models form the “cornerstones” for perturbative and numerical methods in this area. Obtaining exact formulas for the light ''cones'' of an observer in a gravitational field, where hypersurfaces of propagation of massless particles and light in the general case cease to be cone surfaces, also plays an important role for various applied problems, including finding ''retarded'' potentials for electromagnetic radiation of charges moving in a gravitational field, as well as finding ''the retarded time of radiation'' for various astrophysical applications.

Recently, progress has been made both in the development of methods for detecting gravitational waves~\cite{PhysRevLett.116.061102,PhysRevX.9.031040,PhysRevX.11.021053}, and in the development of mathematical methods for their  study of \cite{Christensen2018016903,Osetrin2020Symmetry,Domenech2021398,
Osetrin1455Sym_2023,van_Remortel_2023}.
Based on new mathematical approaches, a number of exact mathematical models of gravitational waves and, in particular, primordial gravitational waves in various cosmological models have been constructed  \cite{OsetrinHomog2006,Osetrin2022894,Osetrin2022EPJP856,Osetrin325205JPA_2023}. Mathematical methods have been developed and a number of exact cosmological models with an electromagnetic field have been constructed, allowing the integration of the equations of motion of charges in the Hamilton-Jacobi formalism  \cite{ObukhovSym14122595,ObukhovUniverse8040245,ObukhovSym15030648,Obukhov10.1063/5.0158054}.

Recent significant results of long-term observations of delays in pulsar signals  \cite{AandArefId0,Reardon_2023,Xu_2023}, which served as one of the motivations for this work, allow us to judge the presence and characteristics of the gravitational wave background of the universe, which in this case can be indirectly detected due to the mentioned effect ''the retarded time of radiation''.

In this work, we will consider obtaining relations for the retarded time of radiation in the class of models of strong gravitational waves without the use of perturbative methods. Moreover, the obtained relations are based on the Hamilton-Jacobi formalism and on obtaining the complete integral for the eikonal equation and will be obtained in a form that allows them to be used both for the field equations of Einstein’s general theory of relativity and for the field equations of modified theories of gravity \cite{Odintsov2007,Odintsov2011, Capozziello2011, Odintsov2017}. This makes it possible, if necessary, to take into account various effects in the models under consideration to describe the phenomena of “dark energy” and “dark matter”.

As a particular example of the application of the obtained general relations for a gravitational wave, the paper considers an exact model of a gravitational wave in the Bianchi universe of type VI  \cite{Osetrin325205JPA_2023}.



\section{Light ''cone'' of an observer in a gravitational wave}

The equation of motion of a test particle of mass $m$ in a gravitational field in the Hamilton-Jacobi formalism in the coordinate system $\{x^\alpha\}$ has the form \cite{LandauEng1}:
\begin{equation}
g^{\alpha\beta}\frac{\partial S}{\partial x^\alpha}\frac{\partial S}{\partial x^\beta}=
m^2c^2
,\qquad
\alpha,\beta,\gamma=0,1,2,3.
\label{HJE}
\end{equation}
Here $g^{\alpha\beta}$ is the space-time metric, $S$ is the action function of the test particle, $c$ is the speed of light. In what follows, to shorten the entries, we will choose the speed of light as unity.

In this work, we use Einstein’s standard rule of summation over repeated subscripts and superscripts, and in places where discrepancies are possible, explicit summation signs are used. We also note that some of the indices used may not correspond to the indices of tensor quantities of space-time, which is always clear from the context, therefore the position of such indices at the bottom or at the top does not have a tensor character, but is determined by the convenience of notation.

The paper considers the model of a ''strong'' gravitational wave with a space-time interval 
of the form~\cite{LandauEng1}:
\begin{equation}
{ds}^2=
2dx^0dx^1+\sum_{p,q} g_{pq}(x^0)\Bigl(dx^p + f^p(x^0) dx^1\Bigr)\Bigl(dx^q + f^q(x^0 ) dx^1\Bigr)
\label{metric1}
,\end{equation}
$$
p,q,r=2,3.
$$
Here $s$ is the space-time interval of the gravitational wave (not to be confused with the particle action function, denoted by the capital letter $S$). The relation (\ref{metric1}) uses a ''privileged'' coordinate system with a wave variable $x^0$, along which the space-time interval vanishes. The metric includes five independent functions of the wave variable $x^0$. Metrics of this type belong to the class of {St{\"{a}}ckel} metrics and in a privileged coordinate system allow integration and obtaining a complete integral for the equation (\ref{HJE}) by the method of separation of variables \cite{Shapovalov1978I,Shapovalov1978II,Shapovalov1979}.

Substitution of the metric (\ref{metric1}) into Einstein's vacuum equations with the cosmological constant
$\Lambda$
\begin{equation}
R_{\alpha\beta}=\Lambda g_{\alpha\beta}
,\end{equation}
where $R_{\alpha\beta}$ is the Ricci tensor,
leads to the following necessary (but not sufficient) conditions for the form of the gravitational wave metric under consideration:
\begin{equation}
f^p(x^0)=0
.\end{equation}

Therefore, further we will consider the metric of a ''strong'' gravitational wave of the form
\begin{equation}
{ds}^2=
2\,dx^0dx^1+\sum_{p,q} g_{pq}(x^0)\,dx^p dx^q
\label{metric2}
,\end{equation}
where there remain three unknown functions of the wave variable $x^0$, which must correspond to the gravitational field equations, which we have not yet explicitly used.

Integrating the Hamilton-Jacobi equation (\ref{HJE}) for the gravitational wave metric (\ref{metric2}) using the separation of variables method, we obtain the following general form of the complete integral for the particle action function $S$ \cite{Osetrin1455Sym_2023}:
\begin{equation}
S(x^\alpha,{\lambda_1} ,{{\lambda_3}} ,\lambda_3)=\phi_0(x^0)+\lambda_k x^k
,\qquad
i,j,k=1,2,3
,\end{equation}
\begin{equation}
\phi_0(x^0)=\frac{1}{2{\lambda_1}}
\Bigl(
m^2x^0-\sum_{p,q} \lambda_p\lambda_q G^{pq}(x^0)
\Bigr)
,\qquad
p,q=2,3
,\end{equation}
\begin{equation}
G^{pq}(x^0)=
\int{g^{pq}(x^0)}\,dx^0
,\qquad
G^{pq}=G^{qp}
\label{functionG}
,\end{equation}
where ${\lambda_1}$, ${{\lambda_3}}$ and $\lambda_3$ are independent constant parameters of the complete integral for the test particle action function, determined by the initial or boundary conditions of the test particle’s motion (initial values of the particle’s momenta). The parameter ${\lambda_1}$ cannot be equal to zero, otherwise the equations and solution degenerate. We also note that solutions to the field equations of the theory of gravitation are not yet used here.

The particle trajectory equations (geodesic line equations) according to the general Hamilton-Jacobi formalism have the form
\begin{equation}
\frac{\partial S}{\partial \lambda_k}=\sigma_k 
\label{traject1}
,\end{equation}
where $\sigma_k$ are new independent constant parameters of the particle trajectory, determined by the initial or boundary conditions of motion of the test particle (coordinates of the initial position of the particle).

The trajectory equations (\ref{traject1}) give:
\begin{equation}
0=-\sigma_1+x^1-\frac{m^2x^0}{2\left({\lambda_1}\right)^2}
+\sum_{p,q} \frac{\lambda_p\lambda_q}{2\left({\lambda_1}\right)^2}\,G^{pq}\bigl(x^0\bigr)
\label{traject1A}
,\end{equation}
\begin{equation}
0=-\sigma_q+x^q
-\sum_p\frac{\lambda_p}{{\lambda_1}}\,G^{pq}\bigl(x^0\bigr)
\label{traject1B}
.\end{equation}
If we consider the variable $x^0$ as a parameter on the trajectory of the particle, then the equation
(\ref{traject1A}) gives $x^1=x^1(x^0)$, and equations (\ref{traject1B}) give $x^q=x^q(x^0)$. Functions $G^{pq}\bigl(x^0\bigr)$ are found from the components of the metric (\ref{functionG}), as solutions to the field equations of some theory of gravity.

The proper time of a particle $\tau$ on its trajectory can be chosen as follows \cite{LandauEng1}:
\begin{equation}
\tau=S\,\Bigl\vert_{m=1}
\label{tau}
.\end{equation}
Using the relations (\ref{traject1A}), (\ref{traject1B}) and (\ref{tau}) we obtain the following form of the trajectory of a test particle of unit mass in a privileged coordinate system:
\begin{equation}
x^0(\tau)={\lambda_1}(\tau-\tau_0)
\label{traject10}
,\end{equation}
\begin{equation}
x^1(\tau)=\sigma_1+\frac{1}{2{\left({\lambda_1}\right)}^2}
\Bigl(
{\lambda_1}(\tau-\tau_0)
-\sum_{p,q} \lambda_p\lambda_q G^{pq}\bigl(x^0(\tau)\bigr)
\Bigr)
\label{traject11}
,\end{equation}
\begin{equation}
x^p(\tau)=\sigma_p+\sum_q\frac{\lambda_q}{{\lambda_1}}\,G^{pq}\bigl(x^0(\tau)\bigr)
\label{traject1p}
.\end{equation}
The constant $\tau_0$ is associated with the presence of an arbitrary additive constant in the complete integral for the particle action.
For a specific test particle, by choosing the origin of coordinates and time, we can choose zero values of the parameters $\tau_0=\sigma_k=0$.

Note that from the equations of motion of the test particle (\ref{traject1A}), (\ref{traject1B})  the relation connecting the variables $x^\alpha$ on the trajectory of the particle follow:
\begin{equation}
\frac{m^2}{{\lambda_1}}x^0-2{\lambda_1}x^1-\lambda_qx^q=
-2{\lambda_1}\sigma_1+{{\lambda_3}}\sigma_2+\lambda_3\sigma_3
=\mbox{const}
.\end{equation}

The motion of massless particles and the trajectories of light rays in a gravitational field are determined by the eikonal equation
\begin{equation}
g^{\alpha\beta}\frac{\partial \Psi}{\partial x^\alpha}\frac{\partial \Psi}{\partial x^\beta}=
0
\label{eikonal}
,\end{equation}
where $\Psi$ is the eikonal function.

Using the method of separation of variables \cite{Shapovalov1978I,Shapovalov1978II,Shapovalov1979} for the eikonal equation (\ref{eikonal}) with the metric (\ref{metric2}) we obtain the complete integral in the form
\begin{equation}
\Psi=\psi_0(x^0)+\sum_i k_i x^i
,\end{equation}
\begin{equation}
\psi_0(x^0)=-\sum_{p,q} \frac{k_pk_q}{2k_1}
\,
  G^{pq}(x^0)
,\end{equation}
where the independent constant parameters $k_i$ are determined by the initial (boundary) values of the radiation wave vector.

In accordance with the Hamilton-Jacobi formalism, we find the trajectories of massless particles and light rays
 in the following form:
\begin{equation}
x^1=\gamma_1-\sum_{p,q} \frac{k_pk_q}{2{\left(k_1\right)}^2}
\,
G^{pq}\bigl(x^0\bigr)
\label{LightEq1}
,\end{equation}
\begin{equation}
x^p=\gamma_p+\sum_q \frac{k_q}{k_1}
\,
G^{pq}\bigl(x^0\bigr)
\label{LightEqp}
,\end{equation}
where the constants $\gamma_i$ are independent parameters of light trajectories, determined by initial or boundary conditions. In the privileged coordinate system used here, the wave variable $x^0$ plays the role of a parameter along the trajectories of massless particles and light rays.

If an observer freely falling in a gravitational wave detects electromagnetic radiation emitted by a charge moving in this gravitational wave, then the radiation propagation trajectory must connect these two world points.

In order to introduce a single synchronized time and spatial variables, it is convenient to move from a privileged coordinate system to a synchronous reference system. To do this, we will use the solution we obtained earlier for the equations of the trajectory of a massive particle (\ref{traject10})-(\ref{traject1p}). If we use independent constant parameters on the particle trajectory $\lambda_k$ as new variables, and the particle’s proper time as a new time variable, then we obtain a synchronous reference system (see \cite{LandauEng1}).

A newly introduced synchronous reference system with variables $y^\alpha=\bigl\{t,y^1,y^2,y^3\bigr\}$, where $t$ is a time variable, and spatial coordinates are specified by the variables $ y^k=\bigl\{y^1,y^2,y^3\bigr\}$, can be constructed by a special transformation of variables obtained from the relations (\ref{traject10})-(\ref{traject1p}) in the following form:
\begin{equation}
x^0 \to {t} y^1
\label{TransX0}
,\end{equation}
\begin{equation}
x^1 \to \frac{{t}}{2 y^1}
-
\sum_{p,q}
\frac{y^py^q}{2{\left(y^1\right)}^2} 
\,
G^{pq}\bigl({t} y^1\bigr)
\label{TransX1}
,\end{equation}
\begin{equation}
x^p\to
\sum_q
\frac{y^q}{2y^1}
\,
G^{pq}\bigl({t} y^1\bigr)
\label{TransXp}
.\end{equation}

In the new synchronous reference frame $y^\alpha=\bigl\{t,y^1,y^2,y^3\bigr\}$ the gravitational wave interval (\ref{metric2}) will take the following form
\begin{equation}
{ds}^2={dt}^2-dl^2={dt}^2+\sum_{i,k}\tilde g{}_{ik}\left(t,y^1,y^2,y^3\right) dy^idy^k
,\quad
i,j,k=1,2,3
,\end{equation}
where $t$ is a single synchronized time,
$dl$ is the spatial distance element, $y^k$ is the spatial variables, $\tilde g{}_{ik}$ is the components of the gravitational wave metric in the new synchronous reference frame (SRF).

Direct calculation of the form of metric components $\tilde g{}_{ik}\left(t,y^1,y^2,y^3\right)$ in the new synchronous  reference frame gives:
\begin{equation}
\tilde g{}_{00} = 1 
,\qquad
\tilde g{}_{0k} = 0 
\label{SinchrMetric0k}
, \end{equation}
\begin{equation}
\tilde g{}_{pq}\left(t,y^1\right)  =
\frac{1}{\left(y^1\right)^2}
\,
{A}_{pq}(ty^1)
,\qquad
p,q,r,s=2,3
.\end{equation}
\begin{equation}
\tilde g{}_{11}\left(t,y^1,y^2,y^3\right) = 
\frac{1}{\left(y^1\right)^4}
\left(
-t^2\left(y^1\right)^2
+\sum_{p,q}y^p y^q {A}_{pq}({t}y^1)
\right)
,\end{equation}
\begin{equation}
\tilde g{}_{1q}\left(t,y^1,y^2,y^3\right)=
-
\sum_{p}
\frac{y^p}{\left(y^1\right)^3}
\,
{A}_{pq}(ty^1)
=
-
\sum_p
\frac{y^p}{y^1}
\,
\tilde g{}_{pq}({t} y^1)
\label{SinchrMetricpq}
,\end{equation}
where the following notations are used
\begin{equation}
{A}_{pq}(ty^1)=
\sum_{r,s=2}^3 
G^{pr}g_{rs}G^{sq}
\label{Fpq}
.\end{equation}
The functions ${A}_{pq}$ are functions of one variable, which, according to the variable transformation used, in the new synchronous reference frame 
 is equal to the product of time $t$ and spatial variable $y^1$.

Thus, the element of spatial distance in the gravitational wave in the new synchronous 
 reference frame 
will take the following form
$$
{dl}^2=
{t}^2\left(\frac{dy^1}{y^1}\right)^2
-
\sum_{p,q}
\tilde g{}_{pq}
\left(
y^p y^q
\left(\frac{{dy^1}}{y^1}\right)^2
-
2\,
{y^p}
\,\frac{{dy^1}}{y^1}{dy^q}
+
{dy^p}{dy^q}
\right)
.$$
We write the above expression in the following form:
\begin{equation}
{dl}^2=
{t}^2\left(\frac{dy^1}{y^1}\right)^2
-
\sum_{p,q}
\tilde g{}_{pq}
\left(
\frac{
y^p
}{{y^1}}
\,
{dy^1}
-
{dy^p}
\right)
\left(
\frac{
y^q
}{{y^1}}
\,
{dy^1}
-
{dy^q}
\right)
.\end{equation}
For convenience, we can introduce a new spatial variable $z$ instead of $y^1$:
\begin{equation}
{z}=\log y^1
,\qquad
y^1=e^{z}
,\end{equation}
then the spatial distance element and the interval in the gravitational wave in the synchronous reference frame will take the following, simpler form
\begin{equation}
{ds}^2={dt}^2-{dl}^2
,\quad
{dl}^2=
{t}^2{d{z}}^2
-
e^{-2z}
{A}_{pq}({t} e^{z})
\Bigl(
y^p d{z}-dy^p
\Bigr)
\Bigl(
y^q d{z}-dy^q
\Bigr)
,\end{equation}
where $t$ is time, the variables $z$, $y^2$ and $y^3$ are spatial coordinates, three functions of one variable ${A}_{pq}({t} e^{z})$ are determined by the field equations of the theory of gravitation.

The equations for the trajectories of test particles (geodesic lines) in a gravitational wave (\ref{traject10})-(\ref{traject1p}) in synchronous  reference frame  will take the form
\begin{equation}
0=-2\sigma_1+\frac{t}{y^1}
-\frac{m^2ty^1}{\left({\lambda_1}\right)^2}
+
\sum_{p,q}
\left(
\frac{\lambda_p\lambda_q}{\left({\lambda_1}\right)^2}
-\frac{y^py^q}{\left(y^1\right)^2}
\right)
\,G^{pq}\!\left(ty^1\right)
,\end{equation}
\begin{equation}
0=-\sigma_q+
\sum_p
\left(
\frac{y^p}{y^1}-\frac{\lambda_p}{{\lambda_1}}
\right)
\,G^{pq}\!\left(ty^1\right)
.\end{equation}
Resolving this system of equations with respect to the spatial coordinates of the particle $y^1(t)$ and $y^p(t)$, we obtain the equations for the trajectory of the test particle in the following form:
\begin{equation}
t^2=\frac{m^2}{\left({\lambda_1}\right)^2} \Bigl(ty^1\Bigr)^2
+2ty^1\! \left(\sigma_1+\frac{\sum_q \lambda_q\sigma_q}{{\lambda_1}}\! \right)
+ty^1\sum_{p,q}\sigma_p\sigma_q\! \left[G^{-1}\left(ty^1\right)\right]^{pq}
\label{eqY1}
,\end{equation}
\begin{equation}
y^p(t)=y^1(t)
\left(
\sum_q
\sigma_q\left[G^{-1}\left(ty^1\right)\right]^{pq}
+\frac{\lambda_p}{{\lambda_1}}
\right)
\label{eqYp}
.\end{equation}
Here the equation (\ref{eqY1}) determines the spatial coordinate $y^1(t)$ of the particle (with initial motion parameters $\lambda_k$, $\sigma_k$) as a function of time t, and the equations (\ref{eqYp}) specify the spatial coordinates $y^p(t)$ of the particle as a function of time t. The functions $\left[G^{-1}\right]^{pq}$ are components of the inverse matrix
for matrix
$G^{pq}\!\left(ty^1(t)\right)$, i.e. are determined by the components of the space-time metric~$g^{pq}$ corresponding to the field equations of a specific theory of gravity.

The equations for the trajectories of light rays in a gravitational wave (\ref{LightEq1})-(\ref{LightEqp}) in synchronous  reference frame  will take the form
\begin{equation}
\frac{t}{y^1(t)}
-
\sum_{p,q}\gamma_p\gamma_q\left[G^{-1}\left(ty^1\right)\right]^{pq}
=
2
\left(
\gamma_1
+\frac{\sum_q k_q\gamma_q}{k_1}
\right)
\label{eqLightY1}
,\end{equation}
\begin{equation}
y^p(t)=
y^1(t)
\left(
\sum_q
\gamma_q\left[G^{-1}\left(ty^1\right)\right]^{pq}
+\frac{k_p}{k_1}
\right)
\label{eqLightYp}
,\end{equation}
where equation (\ref{eqLightY1}) determines the spatial coordinate $y^1(t)$ of the trajectory of a light ray with initial parameters $k_i$, $\gamma_i$ as a function of time t, and equations (\ref{eqLightYp}) define spatial coordinates $y^p(t)$ of the light beam trajectory as a function of time t.

Now we can obtain the relations that determine the hypersurface of the light ''cone'' of an observer moving in a gravitational wave, assuming that this hypersurface is defined by all light rays passing through the world point of the observer $y_o{}^\alpha=\Bigl\{t_o, y_o{}^1,y_o{}^2,y_o{}^3\Bigr\}=
\Bigl\{t_o,y^1(t_o),y^2(t_o),y^3(t_o)\Bigr\}
$. These conditions give us three relations (\ref{eqY1})-(\ref{eqYp}) for six constant parameters $\sigma_i$, $\lambda_i$ of the observer's motion and three relations (\ref{eqLightY1})-(\ref {eqLightYp}) into six parameters $\gamma_i$, $k_i$ for possible light trajectories on the observer's light ''cone''.
Note that the equations for the trajectories of light rays (due to the zero normalization of the wave vector) include the parameters $k_i$ only in the form of combinations $k_p/k_1$, so without loss of generality it is possible to set $k_1=1$.

It is convenient to select the parameters $\lambda_p$, $\sigma_1$ and $k_p$, $\gamma_1$ for fixing on the light ''cone'', then for these parameters we obtain the following values:
\begin{equation}
\lambda_p=
{\lambda_1}
\left(
\frac{{y_o{}^p}}{{y_o{}^1}}
-
\sum_q
\sigma_q\left[G^{-1}\left({t_o}{y_o{}^1}\right)\right]^{pq}
\right)
,\end{equation}
\begin{equation}
\sigma_1=
\frac{1}{2}
\left(
\frac{{t_o}}{{y_o{}^1}}
-\frac{m^2}{\left({\lambda_1}\right)^2}\, {t_o}{y_o{}^1}
+
\sum_{p,q} \sigma_p\sigma_q\left[G^{-1}\left({t_o}{y_o{}^1}\right)\right]^{pq}
\right)
-
\sum_p 
\frac{\sigma_p{y_o{}^p}}{{y_o{}^1}}
,\end{equation}
\begin{equation}
k_p=
{k_1}
\left(
\frac{{y_o{}^p}}{{y_o{}^1}}
-
\sum_q \gamma_q\left[G^{-1}\left({t_o}{y_o{}^1}\right)\right]^{pq}
\right)
\label{LightKp}
,\end{equation}
\begin{equation}
\gamma_1=
\frac{1}{2}
\left(
\frac{{t_o}}{{y_o{}^1}}
+\sum_{p,q} \gamma_p\gamma_q\left[G^{-1}\left({t_o}{y_o{}^1}\right)\right]^{pq}
\right)
-\sum_p\frac{\gamma_p{y_o{}^p}}{{y_o{}^1}}
\label{LightGamma1}
.\end{equation}
Thus, in synchronous  reference frame, all light rays with trajectories (\ref{eqLightY1}), (\ref{eqLightYp}) and trajectory parameters (\ref{LightKp}), (\ref{LightGamma1}) form a light ''cone'' at the world point $y_o{}^\alpha=\Bigl\{t_o,y_o{}^1,y_o{}^2,y_o{}^3\Bigr\}$. Functions $G^{pq}(ty^1)$ defined through the gravitational wave metric (see (\ref{functionG})) and, accordingly, components of the inverse matrix $\left[G^{-1}\right]^ {pq}(ty^1)$ are determined by the explicit form of the metric $g^{pq}$ from the field equations of the theory of gravitation.

Since we did not use field equations for the gravitational wave metric (\ref{metric2}), all the obtained relations can be applied to any theory of gravity, where the movement of test particles and light occurs along geodesic lines of space-time, including Einstein’s general theory of relativity and modified theories of gravity.

\section{
Propagation of radiation from a source ''freely falling'' in a gravitational wave
and
the retarded time of radiation}

Let us now consider an important applied case of application of the obtained relations for the propagation of charge radiation in a gravitational wave. Suppose that when transforming coordinates into a synchronous reference system, we have chosen a system of units, where the charge has unit mass and is at rest with coordinates $y_{ch}^k=\lambda_{(ch)k}$, and the origin of the coordinates and the proper time of the charge (which becomes in synchronous  reference frame  a single synchronized time t) is chosen so that the constants $\sigma_{(ch)k}=0$. Then the equations of motion for this charge in a gravitational wave in synchronous  reference frame  are carried out identically. The observer in the synchronous  reference frame  chosen in this way moves along a trajectory corresponding to the relations (\ref{eqY1})-(\ref{eqYp}) and with eigenvalues of the integration constants $\lambda_k$, $\sigma_k$. If an observer detects the electromagnetic field of a charge at time $t$, then the coordinates of the world point $y_{ch}^\alpha=\Bigl\{t',y_{ch}^k\Bigr\}$ for the charge emitting electromagnetic radiation must lie for the observer on the light ''cone'' of his past ($t>t'$).

The equations for the trajectories of light rays detected by an observer at the world point $\left\{{t_o},{y_o}{}^k\right\}$ in a gravitational wave with a metric (\ref{metric2}) in a synchronous reference frame take the form:
\begin{equation}
\frac{t}{y^1(t)}-\sum_{p,q} \gamma_p\gamma_q\left[G^{-1}\left(ty^1\right)\right]^{pq}
=
2
\left(
\gamma_1
+\frac{\sum_q k_q\gamma_q}{k_1}
\right)
\label{eqLightY1A}
,\end{equation}
\begin{equation}
y^p(t)=
y^1(t)
\left(
\sum_q\gamma_q\left[G^{-1}\left(ty^1\right)\right]^{pq}
+\frac{k_p}{k_1}
\right)
\label{eqLightYpA}
,\end{equation}
where the values of the parameters of light trajectories $k_p$ and $\gamma_1$ are determined by the relations
\begin{equation}
k_p=
{k_1}
\left(
\frac{{y_o{}^p}}{{y_o{}^1}}
-\sum_q\gamma_q\left[G^{-1}\left({t_o}{y_o{}^1}\right)\right]^{pq}
\right)
\label{LightKpA}
.\end{equation}
\begin{equation}
\gamma_1=
\frac{1}{2}
\left(
\frac{{t_o}}{{y_o{}^1}}
+\sum_{p,q} \gamma_p\gamma_q\left[G^{-1}\left({t_o}{y_o{}^1}\right)\right]^{pq}
\right)
-\sum_p\frac{\gamma_p{y_o{}^p}}{{y_o{}^1}}
\label{LightGamma1A}
.\end{equation}
For the observer coordinates $\Bigl\{t,y_o{}^1,y_o{}^2,y_o{}^3\Bigr\}$ we have:
\begin{equation}
t^2=\frac{m^2}{\left({\lambda_1}\right)^2} \Bigl(ty_o{}^1\Bigr)^2\!
+2ty_o{}^1\! \left(\!\sigma_1+\frac{\sum_q \lambda_q\sigma_q}{{\lambda_1}}\!\right)
+ty_o{}^1\!\sum_{p,q} \sigma_p\sigma_q\left[G^{-1}\!\!\left({t}{y_o{}^1}\right)\right]^{pq}
\label{eqY1A}
,\end{equation}
\begin{equation}
y_o{}^p(t)=y_o{}^1(t)
\left(
\sum_q
\sigma_q\left[G^{-1}\left({t_o}{y_o{}^1}\right)\right]^{pq}
+\frac{\lambda_p}{{\lambda_1}}
\right)
\label{eqYpA}
.\end{equation}
In this case, along the path of propagation of the light signal from the emitting charge to the observer, the following condition must be satisfied (the speed of light $c$ is chosen as unity):
$$
c(t-t')=t-t'=
\int_{t'}^t\,dl
=
$$
\begin{equation}
=
\int_{t'}^t\!\! dt {\sqrt{
\frac{{t}^2}{\left(y^1\right)^{\! 2}}
\biggl(\!\frac{dy^1}{dt}\!\biggr)^2
-
\tilde g{}_{pq}\!
\left(\!
\frac{y^p y^q}{\left(y^1\right)^{\! 2}}
\biggl(\!\frac{{dy^1}}{dt}\!\biggr)^{\! 2}
\!
-
\left(
2\,
\frac{y^p}{y^1}
\,\frac{{dy^1}}{dt}
-
\frac{dy^p}{dt}
\right)\!
\frac{dy^q}{dt}
\!\right)
}}
\label{SatisfEq}
,\end{equation}
where $t'$ is the moment of emission of radiation by the charge, $t$ is the moment of detection of radiation by the observer, and the integral is taken along the trajectory of the light signal (\ref{eqLightY1A})-(\ref{eqLightYpA}), components of the metric $\tilde g{ }_{pq}$ of a gravitational wave in synchronous  reference frame  takes the form:
\begin{equation}
\tilde g{}_{pq} =
\frac{1}{\left(y^1\right)^2}
\,
\sum_{r,s=2}^3
G^{pr}
g_{rs}
G^{sq}
,\end{equation}

Let's find the derivatives $dy^1/dt$, $dy^p/dt$
  from the relations (\ref{eqLightY1A})-(\ref{eqLightYpA}).
In this case, we will take into account that the matrices satisfy the following relation:
\begin{equation}
-\left[\!\frac{d G^{-1}}{dt}\!\right]^{pq}=
\left[\! G^{-1}\frac{dG}{dt}G^{-1}\!\right] ^{pq}
=
\left(\!
y^1+t\,\frac{dy^1}{dt}
\!\right)\!
\sum_{r,s=2}^3[G^{-1}]^{pr}g^ {rs}[G^{-1}]^{sq}
.\end{equation}
Differentiating the relations (\ref{eqLightY1A})-(\ref{eqLightYpA}) with respect to time $t$ and solving the resulting equations for $dy^1/dt$, $dy^p/dt$, we find for the derivatives of the coordinates on the light trajectories rays:
\begin{equation}
\frac{dy^1}{dt}=
\left(\frac{y^1(t)}{t}\right)
\frac{1+\left(y^1(t)\right)^2\sum_{p,q} \gamma_p\gamma_q\,{B}^{pq}\!\left(ty^1(t)\right)
}{
1-\left(y^1(t)\right)^2\sum_{p,q} \gamma_p\gamma_q\,{B}^{pq}\!\left(ty^1(t)\right)
}
\label{DifY1}
,\end{equation}
$$
\frac{dy^p}{dt}=
\frac{dy^1}{dt}
\left(
-
ty^1(t)\sum_q\gamma_q\,{B}^{pq}\!\left(ty^1(t)\right)
+
\sum_q
\gamma_q\left[G^{-1}\left(ty^1\right)\right]^{pq}
+\frac{k_p}{k_1}
\right)
$$
\begin{equation}
\mbox{}
-
\left(y^1(t)\right)^2\sum_q\gamma_q\,{B}^{pq}\!\left(ty^1(t)\right)
\label{DifYp}
,\end{equation}
where additional notation ${B}^{pq}$ is used for the components of the matrix of the following form:
\begin{equation}
{B}^{pq}\!\left(ty^1(t)\right)=
\sum_{r,s=2}^3[G^{-1}]^{pr}g^{rs}[G^{-1}]^{sq}
\label{matrixBpq}
,\end{equation}
\begin{equation}
{B}^{22}=
\frac{
g^{22} {G^{33}}^2-2 g^{23} G^{23} G^{33}+g^{33}{G^{23}}^2 
}{\left({G^{23}}^2-G^{22} G^{33}\right)^2}
,\end{equation}
\begin{equation}
{B}^{23}={B}^{32}=
\frac{
g^{23} \left(G^{22} G^{33}+{G^{23}}^2\right)
-G^{23} (g^{22} G^{33}+G^{22} g^{33})
}{\left({G^{23}}^2-G^{22} G^{33}\right)^2}
,\end{equation}
\begin{equation}
{B}^{33}=
\frac{
g^{22} {G^{23}}^2 -2 g^{23} G^{22}  G^{23} +g^{33} {G^{22}}^2 
}{\left({G^{23}}^2-G^{22} G^{33}\right)^2}
,\end{equation}
\begin{equation}
G^{pq}(x^0)=
\int{g^{pq}(x^0)}\,dx^0
,\qquad
G^{pq}=G^{qp}
.\end{equation}
Note that the product of matrices
${A}_{pq}$
(\ref{Fpq})
and ${B}^{pq}$ (\ref{matrixBpq}) gives
\begin{equation}
\sum_{r=2}^3
{A}_{pr}
{B}^{rq}=
\delta_p^q
,\end{equation}
where $\delta_p^q$ is the identity matrix.

Along the trajectory of light rays in a gravitational wave we obtain
$$
\left(\frac{dl}{dt}\right)^2=
\sum_{i,j}
\tilde g{}_{ij}
\frac{dy^i}{dt}
\frac{dy^j}{dt}
=
\left(\frac{t}{y^1}\right)^2
\left(\frac{dy^1}{dt}\right)^2
$$
$$
\mbox{}
-
\sum_{p,q} \frac{{A}_{pq}}{\left(y^1\right)^4}
\left(
y^1\frac{dy^p}{dt}-y^p\frac{dy^1}{dt}
\right)
\left(
y^1\frac{dy^q}{dt}-y^q\frac{dy^1}{dt}
\right)
=
$$
\begin{equation}
\mbox{}
=
\left(\frac{t}{y^1}\right)^2
\left(\frac{dy^1}{dt}\right)^2
-
\sum_{p,q} 
{A}_{pq}\,
\frac{d}{dt}
\left(
\frac{y^p}{y^1}
\right)
\frac{d}{dt}
\left(
\frac{y^q}{y^1}
\right)
=1
,\end{equation}
where the relations (\ref{DifY1}), (\ref{DifYp}) (or (\ref{eqLightYpA})) were used), the speed of light $c$ was previously chosen by us to be unity.

Thus, along the trajectory of light rays, the condition (\ref{SatisfEq}) is satisfied, and the spatial components of the speed of light propagation in a gravitational wave have the form (\ref{DifY1})-(\ref{DifYp}).

Let us find the connection between the moment of emission of the charge $t'$ and the coordinates and time of the observer. World point of the radiating charge with coordinates $\left\{ t',{\lambda_1}, {{\lambda_3}},\lambda_3 \right\}$ and world point $\left\{t,y^1,y^2,y^3\right \}$  of the observer detecting radiation,
  are connected by the equations of the trajectory of light. 
\begin{equation}
\frac{{t'}}{{{\lambda_1}}}
-\sum_{p,q} \gamma_p\gamma_q\left[G^{-1}\left({{\lambda_1}}{t'}\right)\right]^{pq}
=
2
\left(
\gamma_1
+\frac{\sum_q k_q\gamma_q}{k_1}
\right)
\label{eqLightY1Charge}
,\end{equation}
\begin{equation}
{\lambda_p}=
{{\lambda_1}}
\left(
\sum_q
\gamma_q\left[G^{-1}\left({{\lambda_1}}{t'}\right)\right]^{pq}
+\frac{k_p}{k_1}
\right)
\label{eqLightYpCharge}
,\end{equation}
 
\begin{equation}
\frac{t}{y^1(t)}-
\sum_{p,q} \gamma_p\gamma_q\left[G^{-1}\left(ty^1\right)\right]^{pq}
=
2
\left(
\gamma_1
+\frac{\sum_q k_q\gamma_q}{k_1}
\right)
\label{eqLightY1Observer}
,\end{equation}
\begin{equation}
y^p(t)=
y^1(t)
\left(
\sum_q
\gamma_q\left[G^{-1}\left(ty^1\right)\right]^{pq}
+\frac{k_p}{k_1}
\right)
\label{eqLightYpObserver}
.\end{equation}

Eliminating the radiation trajectory parameters $k_i$ and $\gamma_1$ from the equations (\ref{eqLightY1Charge})-(\ref{eqLightYpObserver}), we obtain relations for $t'$, $t$, $y^1(t)$ and $y^p(t)$ of the following form:
\begin{equation}
\sum_q
\gamma_q
\Bigl[
G^{-1}\left(ty^1\right)
-
G^{-1}\left({{\lambda_1}}{t'}\right)
\Bigr]^{pq}
=
\frac{y^p(t)}{y^1(t)}
-
\frac{\lambda_p}{{\lambda_1}}
\label{EqForGammaP}
,\end{equation}


\begin{equation}
\sum_p
\gamma_p
\left(
\frac{y^p(t)}{y^1(t)}
-
\frac{\lambda_p}{{\lambda_1}}
\right)
=
\frac{t}{y^1(t)}
-
\frac{{t'}}{{{\lambda_1}}}
\label{EqConnection}
.\end{equation}
Here the equations (\ref{EqForGammaP}) are used to determine the parameters of the radiation trajectory $\gamma_p$. Since the observer's position is parameterized by time $t$, then 
 the parameters $\gamma_p$ in the relations (\ref{EqForGammaP})-(\ref{EqConnection}) formally are functions of $t$ and $t'$. 
Substitution of the parameters $\gamma_p$ as solutions of the equations (\ref{EqForGammaP}) into the equation (\ref{EqConnection}) gives a relation connecting the time instant $t'$ of charge radiation and the observer's variables when detecting this radiation - the instant $t$ and spatial coordinates $y^1(t)$ and $y^p(t)$.
The $\lambda_k$ parameters are determined by the initial conditions for the charge and specify the initial position of the charge in the synchronous reference frame. The functions $G^{pq}$ are defined in terms of the gravitational wave metric $g^{pg}$ by the relations
(\ref{functionG}). The explicit form of the metric $g^{pg}$ is refined from the gravitational field equations.

From equations (\ref{EqForGammaP}) for the radiation parameters $\gamma_p$ we obtain the following form:
\begin{equation}
\gamma_p=
\sum_q
\left[
\Bigl(
G^{-1}\left(ty^1\right)
-
G^{-1}\left({{\lambda_1}}{t'}\right)
\Bigr)^{-1}
\right]_{pq}
\left(
\frac{y^q(t)}{y^1(t)}
-
\frac{\lambda_q}{{\lambda_1}}
\right)
\label{gammaP}
,\end{equation}

We can introduce new functions $z^k(t)$ instead of the spatial coordinates of the observer $y^k(t)$:
\begin{equation}
z^k(t)=ty^k(t)
,\qquad
k=1,2,3
.\end{equation}
Then the equations 
(\ref{EqConnection})-(\ref{gammaP}) give the following
relation for ''the retarded time'', connecting the emission time $t'$ of the source and the signal detection time $t$ by the observer:
\begin{equation}
t^2=F\left(t',z^k(t)\right)
\label{functionF}
,\end{equation}
where the function $F\left(t',z^k(t)\right)$ is determined from (\ref{EqForGammaP})-(\ref{EqConnection}) by the following relation:
\begin{equation}
F
=
{z^1}\! 
\left(\! 
\frac{{t'}}{{{\lambda_1}}}
+\! 
\sum_{p,q}
\left[
\Bigl(
G^{-1}\!\left(z^1\right)
\! - \!
G^{-1}\!\left({{\lambda_1}}{t'}\right)
\Bigr)^{-1}
\right]_{pq}
\! 
\left(\! 
\frac{z^p}{z^1}
\! - \!
\frac{\lambda_p}{{\lambda_1}}
\! \right)
\left(\! 
\frac{z^q}{z^1}
\! - \!
\frac{\lambda_q}{{\lambda_1}}
\! \right)
\! \right)
\label{EqConnection2}
.\end{equation}
Here $t'$ 
is the moment of emission of radiation by the source, 
 and $t$ is the moment of radiation detection by the observer.
We have excluded parameters of the light signal $k_i$ and $\gamma_i$ from the relations (\ref{EqForGammaP})-(\ref{EqConnection}).
The parameters $\lambda_k$ are determined by the position of the charge in the synchronous reference frame.

If the observer freely ''falls'' in the field of a gravitational wave, then his coordinates correspond to the equations of motion (\ref{eqY1})-(\ref{eqYp}). From the equations (\ref{eqYp}) we can find for the observer the form of the functions $z^p(t)$. 
Substituting them into the equation (\ref{EqConnection2}), we obtain
refined relation for ''the retarded time'':
\begin{equation}
t^2=
P\left(t',z^1(t)\right)
\label{t2P}
,\end{equation}
$$
P\left(t',z^1\right)
=
{z^1(t)}
\,
\Biggl(
\frac{{t'}}{{{\lambda_1}}}
+
\sum_{p,q,r,s}
\left[
\Bigl(
G^{-1}\!\left(z^1\right)
-
G^{-1}\!\left({{\lambda_1}}{t'}\right)
\Bigr)^{-1}
\right]_{pq}
\ast
$$
\begin{equation}
\ast
\Bigl(
\left[G^{-1}\left(z^1\right)\right]^{pr}
{\tilde\sigma{}_r}
+
\delta_p
\Bigr)
\Bigl(
\left[G^{-1}\left(z^1\right)\right]^{qs}
{\tilde\sigma{}_s}
+
\delta_q
\Bigr)
\Biggr)
\label{EqConnection3}
,\end{equation}
\begin{equation}
\delta_p=
{\tilde\lambda{}_p}/{\tilde\lambda{}_1}
-
{\lambda_p}/{{\lambda_1}}
,\end{equation}
where the parameters ${\tilde\lambda{}_k}$ and $\tilde\sigma{}_p$ are determined from the initial conditions of the observer's motion, the parameters $\delta_p$ are determined through the spatial coordinates of the charge ${\lambda_k}$ and the observer's motion parameters ${\tilde\lambda{}_k}$.

The quantities $z^1=ty^1(t)$, where $y^1$ is the spatial coordinate of an observer freely ''falling'' in a gravitational wave, correspond to the equation (\ref{eqY1}), which can be written in the following form:
\begin{equation}
\frac{t^2}{z^1(t) }
=
\frac{m^2}{\tilde\lambda{}_1{}^2}\,z^1(t)
+\sum_{p,q} \tilde\sigma{}_p\tilde\sigma{}_q\left[G^{-1}\left(z^1\right)\right]^{pq}
+2 \left(\tilde\sigma{}_1+\frac{\sum_q \tilde\lambda{}_q\tilde\sigma{}_q}{\tilde\lambda{}_1}\right)
\label{eqY1P}
.\end{equation}

Finally, if we consider the relationship between the time intervals at the source $dt'$ and corresponding time interval at the observer $dt$,
 then by differentiating the relation (\ref{t2P}) for ''the retarded time'', we obtain the following relation for the time intervals of the radiation source and detector (observer) in a gravitational wave:
\begin{equation}
\left(
2t
-
\frac{\partial P}{\partial z^1}\frac{d z^1}{dt}
\right)
dt=
\frac{\partial P}{\partial t'}\,dt'
\quad
\to
\quad
dt=\frac{
{\partial P}/{\partial t'}
}{
2t-\left({\partial P}/{\partial z^1}\right)\left({d z^1}/{dt}\right)
}
\,
dt'
\label{DeltaT}
,\end{equation}
where we obtain from the relation (\ref{EqConnection3}) 
$$
\frac{\partial P}{\partial z^1}=
\frac{P\left(t',z^1\right)}{z^1}
+ z^1
\sum_{p,q,r,s}
\Biggl(
C_{pq}(t',z^1)
\Bigl(
\left[G^{-1}\left(z^1\right)\right]^{pr}
\tilde\sigma{}_r
+
\delta_p
\Bigr)
$$
\begin{equation}
\mbox{}
+2\,
\left[
\Bigl(
G^{-1}\!\left(z^1\right)
-
G^{-1}\!\left({{\lambda_1}}{t'}\right)
\Bigr)^{-1}
\right]_{pq}
\left[
\frac{dG^{-1}\! \left(z^1\right)}{dz^1}\right]^{pr}
\! \! \tilde\sigma{}_r
\Biggr)
\Bigl(
\! \left[G^{-1}\! \left(z^1\right)\right]^{qs}
\tilde\sigma{}_s
+
\delta_q
\Bigr)
\label{dPdz1}
,\end{equation}
where matrix $C_{pq}$ is the product of three matrices 
($I$ is the identity matrix):
$$
C_{pq}(t',z^1(t))=
-
\sum_{r,s}
\left[
\Bigl(
I-G(z^1)\, G^{-1}({\lambda_1}t')
\Bigr)^{-1}
\right]_{pr}
g^{rs}(z^1)
*
$$
\begin{equation}
*
\left[
\Bigl(
I-G^{-1}({\lambda_1}t')\, G(z^1)
\Bigr)^{-1}
\right]_{sq}
.\end{equation}
Similarly, from the relation (\ref{EqConnection3}) we obtain
\begin{equation}
\frac{\partial P}{\partial t'}=
\frac{z^1}{{\lambda_1}}
-
z^1\! \! 
\sum_{p,q,r,s}
\! 
D_{pq}(t',z^1)
\Bigl(
\! 
\left[G^{-1}\! \left(z^1\right)\right]^{pr}
\! \! 
\tilde\sigma{}_r\!
+
\delta_p
\Bigr)
\! 
\Bigl(
\! 
 \left[G^{-1}\! \left(z^1\right)\right]^{qs}
\! \! 
\tilde\sigma{}_s\!
+
\delta_q
\Bigr)
\label{dPdt}
,\end{equation}
where matrix $D_{pq}$ is the product of three matrices
($I$ is the identity matrix):
$$
D_{pq}(t',z^1(t))=
-
{\lambda_1}
\sum_{r,s}
\left[
\Bigl(
I-G({\lambda_1}t')\, G^{-1}(z^1)
\Bigr)^{-1}
\right]_{pr}
g^{rs}({\lambda_1}t')
*
$$
\begin{equation}
*
\left[
\Bigl(
I-G^{-1}(z^1)\, G({\lambda_1}t')
\Bigr)^{-1}
\right]_{sq}
.\end{equation}
From the equation of motion of the observer (\ref{eqY1P}) in a gravitational wave we obtain
\begin{equation}
\frac{d z^1}{dt}=
\frac{2t}{z^1}
\left(
\frac{t^2}{(z^1)^2}
+\frac{m^2}{\tilde\lambda{}_1{}^2}
+
\sum_{p,q} 
\tilde\sigma{}_p\tilde\sigma{}_q\left[\frac{d G^{-1}\left(z^1\right)}{dz^1}\right]^{pq}
\right)^{-1}
\label{dz1dt}
.\end{equation}
Let us also recall that the components of matrices $G(x^0)$ satisfy the following relations:
\begin{equation}
G^{pq}(x^0)=\! \int{\! g^{pq}(x^0)}\,dx^0
,\quad
\left[ \frac{d G^{-1}(x^0)}{dx^0} \right]^{pq}\! =
-\sum_{r,s}\left[G^{-1}\right]^{pr}\!  g^{rs}\left[G^{-1}\right]^{rq}
,\end{equation}
where $g^{pq}(x^0)$ are the components of the gravitational wave metric (\ref{metric2}).

Thus, the relations (\ref{DeltaT}), (\ref{dPdz1}), (\ref{dPdt}) and (\ref{dz1dt}) define the general form of the connection between the time intervals of the radiation source and the detector (observer), freely
''falling''  in a gravitational wave.

The resulting relationships can be used, among other things, to estimate and compare the periods of radiation of various periodic astrophysical sources of radiation in gravitational waves such as pulsars.

  Since we did not explicitly use the equations of the gravitational field, the general relations obtained above are valid for all metric theories of gravity, where the 
 motion of test particles and radiation occurs along geodesic lines of space-time.

\section{
The retarded time for radiation in a gravitational wave of the Bianchi type VI universe
}

As an example of the application of the general relations obtained in the work, let us consider the case of a gravitational wave for the Bianchi type VI universe, the metric of which can be represented in a privileged coordinate system using the wave variable $x^0$ in the following form \cite{Osetrin325205JPA_2023}:
$$
{ds}^2=2\,dx^0dx^1-\frac{1}{\sin^2{\theta}}
\Bigl(
\left({x^0}\right)^{1+\sin{\theta}}{dx^2}^2
$$
\begin{equation}
\mbox{}
+2\,x^0\! \cos{\left(\theta\right)}\, dx^2dx^3
+\left({x^0}\right)^{1-\sin{\theta}}{dx^3}^2
\Bigr)
\label{metricBianchiVI}
,\end{equation}
\begin{equation}
0<\theta<\pi
,\end{equation}
where $x^0$ is the wave variable along which the space-time interval vanishes, the constant $\theta$ is the angular parameter of the gravitational wave model for the Bianchi type VI universe.
The metric (\ref{metricBianchiVI}) is an exact solution of Einstein's vacuum equations for a strong gravitational wave \cite{Osetrin325205JPA_2023} and has the symmetries of the homogeneous but non-isotropic Bianchi type VI model of the universe.

Non-zero components of the Riemann curvature tensor $R_{\alpha\beta\gamma\delta}$ for the metric (\ref{metricBianchiVI}) in the privileged coordinate system have the form:
\begin{equation}
{R}_{0202} = -\frac{1}{2} \cot ^2(\theta )\, \left(x^0\right)^{\left(-1+\sin \theta \right)}
,\end{equation}
\begin{equation}
{R}_{0302} = -\frac{\cot (\theta ) \csc (\theta )}{2 {x^0}}
,\end{equation}
\begin{equation}
{R}_{0303} = -\frac{1}{2} \cot ^2(\theta )\, \left(x^0\right)^{\left(-1-\sin \theta \right)}
.\end{equation}

The spacetimes (\ref{metricBianchiVI}) is plane-wave since it admits a 
 constant vector~$K^{\alpha}$:
\begin{equation}
\nabla_{\beta} K_{\alpha}=0
\quad
\to
\quad
K^{\alpha}=\bigl( 0,1,0,0 \bigl)
.
\end{equation}
This vector specifies the direction of propagation of a plane gravitational wave.

The gravitational wave metric (\ref{metricBianchiVI}) has a subgroup of motions that specifies the homogeneity of this model based on the following Killing vectors $X_{(1)}$, $X_{(2)}$, and $X_{(3)}$ :
\begin{equation}
X^{\alpha}_{(1)}=\bigl(0,0,1,0\bigr),
\quad
X^{\alpha}_{(2)}=\bigl(0,0,0,1\bigr),
\quad
X^{\alpha}_{(3)}=\bigl(-x^0,\, x^1,\, {p} x^2,\, {q} x^3\bigr)
.\end{equation}
An additional null Killing vector has the form
\begin{equation}
X^{\alpha}_{(0)}=K^{\alpha}=\bigl(0,1,0,0\bigr)
,
\qquad
g_{{\alpha}{\beta}}X^{\alpha}_{(0)}X^{\beta}_{(0)}=0
.
\end{equation}
These Killing vectors correspond to the following commutation relations
\begin{equation}
\left[X_{(0)},X_{(1)}\right]=0
,\qquad
\left[X_{(0)},X_{(2)}\right]=0
,\qquad
\left[X_{(0)},X_{(3)}\right]=X_{(0)}
,\qquad
\end{equation}
\begin{equation}
\left[X_{(1)},X_{(2)}\right]=0
,\qquad
\left[X_{(1)},X_{(3)}\right]=p\, X_{(1)}
,\qquad
\left[X_{(2)},X_{(3)}\right]=q\, X_{(2)}
.\end{equation}
For the model under consideration, it is convenient to use a constant angular parameter ${\theta}$:
\begin{equation}
p=
\frac{1+\sin\theta}{2}
,\qquad
q=
\frac{1-\sin\theta}{2}
,\qquad
0<\theta<\pi
.\end{equation}
The three-dimensional homogeneity subgroup of the type VI Bianchi universe of the model under consideration is formed by the Killing vectors $X_{(1)}$, $X_{(2)}$ and $X_{ (3)}$.

Transition $x^\alpha \to y^\alpha=\left({t},y^1,y^2,y^3 \right)$ from a privileged coordinate system to a synchronous reference system (where ${t}$ is time) according to the relations (\ref{TransX0})-(\ref{TransXp}) takes the following form
\begin{equation}
x^0 =  {{t}}{y^1}  
, \end{equation}
\begin{equation}
x^1\!  =  \! 
\frac{
{{t}} {y^1}\! 
+
\csc (\theta ) \! 
\left(
\! 
{y^3}^2 ({{t}} {y^1})^{\sin \theta } 
-{y^2}^2 ({{t}} {y^1})^{-\sin \theta }
\! 
\right)
\! 
-\! 2 {y^2} {y^3}\!  \cos (\theta ) \log ({{t}} {y^1})
}{2 {y^1}^2} 
, \end{equation}
\begin{equation}
x^2 =  \frac{{y^2} \csc (\theta ) ({{t}} {y^1})^{-\sin (\theta )}+{y^3} \cos (\theta ) \log ({{t}} {y^1})}{{y^1}} 
, \end{equation}
\begin{equation}
x^3 =  \frac{{y^2} \cos (\theta ) \log ({{t}} {y^1})-{y^3} \csc (\theta ) ({{t}} {y^1})^{\sin (\theta )}}{{y^1}} 
, \end{equation}

Gravitational wave metric (\ref{metricBianchiVI}) in the Bianchi type VI universe in the synchronous reference frame $y^\alpha=\left\{{t},y^1,y^2,y^3\right\}$ in in accordance with the relations (\ref{SinchrMetric0k})-(\ref{SinchrMetricpq}) will take the following form:
\begin{equation}
{\tilde g}{}_{00} = 1 
,\qquad
{\tilde g}{}_{01} = 0 
,\qquad
{\tilde g}{}_{02} = 0 
,\qquad
{\tilde g}{}_{03} = 0 
, \end{equation}
$$
{\tilde g}{}_{11}({t},y^k) =
 -\frac{
 {{t}}
 }{{y^1}^3 ({{t}} {y^1})^{\sin (\theta )} } 
\biggl[
 \csc ^4(\theta ) \left({y^3}^2 ({{t}} {y^1})^{2 \sin (\theta )}+{y^2}^2\right)
$$
$$
\mbox{}
 +2 \csc ^3(\theta ) \left({y^2}^2-{y^3}^2 ({{t}} {y^1})^{2 \sin (\theta )}\right)
$$
$$
\mbox{}
 +\csc ^2(\theta ) \left({y^3}^2 ({{t}} {y^1})^{2 \sin (\theta )}+{y^2}^2\right)
 -2 \csc (\theta ) \left({y^2}^2-{y^3}^2 ({{t}} {y^1})^{2 \sin (\theta )}\right)
$$
$$
\mbox{}
 +\cot ^2(\theta ) (\log ({{t}} {y^1})-1) 
\Bigr(
 (\log ({{t}} {y^1})+1) \left({y^3}^2 ({{t}} {y^1})^{2 \sin (\theta )}+{y^2}^2\right)
$$
$$
\mbox{}
 +2 \csc (\theta ) \left({y^2}^2-{y^3}^2 ({{t}} {y^1})^{2 \sin (\theta )}\right)
\Bigl)
$$
$$
\mbox{}
 -\frac{1}{4} {y^2} {y^3} \cot (\theta ) \csc ^3(\theta ) ({{t}} {y^1})^{\sin (\theta )} 
\Bigl(
 \cos (4 \theta ) \log ^2({{t}} {y^1})-\log ^2({{t}} {y^1})+8
\Bigr)
$$
\begin{equation}
\mbox{}
 -{y^3}^2 ({{t}} {y^1})^{2 \sin (\theta )}+({{t}} {y^1})^{\sin (\theta )+1}-{y^2}^2
\biggr]
, \end{equation}
$$
{\tilde g}{}_{12}({t},y^k) = 
\frac{{y^2} {{t}} ({{t}} {y^1})^{-\sin (\theta )} 
}{{y^1}^2}
\Bigl(
\cot ^2(\theta ) \csc (\theta ) \left(2 \log ({{t}} {y^1})-1\right)
$$
$$
\mbox{}
+\cot ^2(\theta ) \log ^2({{t}} {y^1})
+\csc ^4(\theta )+\csc ^3(\theta )-\csc (\theta )
\Bigr)
$$
$$
\mbox{}
+
\frac{
{{t}}{y^3}  
}{{y^1}^2} 
\Bigl(
\cos (\theta ) \log ({{t}} {y^1}) 
\left(\cot ^2(\theta ) \left(\log ({{t}} {y^1})-1\right)-1\right)
$$
\begin{equation}
+\cot (\theta ) \csc (\theta ) \log ({{t}} {y^1}) 
-\cot (\theta ) \csc ^3(\theta ) 
\Bigr)
,\end{equation}
$$
{\tilde g}{}_{13}({t},y^k) = 
\frac{{{t}} {y^2} 
}{{y^1}^2}
\Bigl(
\cos (\theta ) \cot ^2(\theta ) \log ^2({{t}} {y^1})-\cot (\theta ) \csc ^3(\theta )
\Bigr)
$$
\begin{equation}
\mbox{}
+\frac{
{{t}}  {y^3} 
({{t}} {y^1})^{\sin (\theta )}
}{{y^1}^2} 
\Bigr(
\csc ^4(\theta ) 
+\cot ^2(\theta ) \log ^2({{t}} {y^1}) 
-2 \cot ^2(\theta ) \csc (\theta ) \log ({{t}} {y^1}) 
\Bigr)
, \end{equation}
\begin{equation}
{\tilde g}{}_{22}({t},y^1) = -\frac{
{{t}} \csc ^2(\theta ) ({{t}} {y^1})^{-\sin (\theta )} 
}{{y^1}}
\Bigl(
\csc ^2(\theta )
$$
$$
\mbox{}
+
\cos (\theta ) \cot (\theta ) \log ({{t}} {y^1}) \left(\sin (\theta ) \log ({{t}} {y^1})+2\right)
\Bigr)
, \end{equation}
\begin{equation}
{\tilde g}{}_{23}({t},y^1) = \frac{
{{t}} \cot (\theta ) 
}{{y^1}} 
\Bigl(
\csc ^3(\theta )-\cos (\theta ) \cot (\theta ) \log ^2({{t}} {y^1})
\Bigr)
,\end{equation}
\begin{equation}
{\tilde g}{}_{33}({t},y^1) =
 -\frac{
 {{t}} \csc ^4(\theta ) ({{t}} {y^1})^{\sin (\theta )} 
  }{{y^1}} 
\Bigl(
1
 $$
$$
\mbox{}
+
\sin (\theta ) \cos ^2(\theta ) \log ({{t}} {y^1}) \left(\sin (\theta ) \log ({{t}} {y^1})-2\right)
 \Bigr)
, \end{equation}
where ${t}$ is a time variable, $y^k=\left\{y^1,y^2,y^3\right\}$ are spatial variables, the angular parameter $\theta$ is a parameter of the Bianchi VI type universe.

For the matrix $G$ in the relation (\ref{functionG}) we obtain the following form:
\begin{equation}
G(x^0)=
\left(
\begin{array}{cc}
 \csc (\theta ) {x^0}^{-\sin (\theta )} & \cos (\theta ) \log ({x^0}) \\
 \cos (\theta ) \log ({x^0}) & -\csc (\theta ) {x^0}^{\sin (\theta )} \\
\end{array}
\right)
.\end{equation}
Then the inverse matrix $G^{-1}$ will take the following form:
\begin{equation}
G^{-1}(x^0)=
\left(
\begin{array}{cc}
 \frac{\sin (\theta ) {x^0}^{\sin (\theta )}}{\sin ^2(\theta ) \cos ^2(\theta ) \log ^2({x^0})+1} &
   \frac{{\log ({x^0})}}{\cos (\theta ) \log^2 ({x^0})+\csc ^2(\theta ) \sec (\theta )} \\
    \frac{{\log ({x^0})}}{\cos (\theta ) \log^2 ({x^0})+\csc ^2(\theta ) \sec (\theta )}
 & -\frac{\sin
   (\theta ) {x^0}^{-\sin (\theta )}}{\sin ^2(\theta ) \cos ^2(\theta ) \log ^2({x^0})+1} \\
\end{array}
\right)
.\end{equation}

For the matrix components $B^{pq}$ in the relation (\ref{matrixBpq}) we obtain:
\begin{equation}
B^{22} (x^0)=
-
\frac{\sin ^2 (\theta )  \left(\sin (\theta ) \cos ^2(\theta ) \log ({x^0}) \left(\sin (\theta ) \log ({x^0})-2\right)+1\right)}{{x^0}^{1-\sin (\theta )}\left(\sin ^2(\theta ) \cos^2(\theta ) \log ^2({x^0})+1\right)^2} 
,\end{equation}
\begin{equation}
B^{23} (x^0)=B^{32} (x^0)=
\frac{\cos ^3(\theta ) \log ^2({x^0})-\cot (\theta ) \csc (\theta )
}{ {x^0} \left(\cos ^2(\theta ) \log ^2({x^0})+\csc   ^2(\theta )\right)^2} 
,\end{equation}
\begin{equation}
B^{33} (x^0)=
-
\frac{\sin ^2(\theta )  
\left(\sin (\theta ) \cos ^2(\theta ) 
 \log ({x^0}) \left(\sin (\theta ) \log ({x^0})+2\right)+1\right)}{
 {x^0}^{1+\sin (\theta )}
 \left(\sin ^2(\theta ) \cos ^2(\theta ) \log
   ^2({x^0})+1\right)^2} 
.\end{equation}

The spatial coordinate $y^1(t)$ along the radiation trajectory as a function of time is determined by the equation (\ref{eqLightY1}), which takes the following form
\begin{equation}
0=
\frac{\sin (\theta )
\left(
{\gamma_3}^2\left({ty^1}\right)^{-\sin (\theta )}
-{\gamma_2}^2 \left({ty^1}\right)^{\sin (\theta )}
\right)
}{\sin ^2(\theta ) \cos ^2(\theta ) \log^2({ty^1})+1}
$$
$$
\mbox{}
-\frac{2 {\gamma_2} {\gamma_3} \cos (\theta ) \log ({ty^1})}{\cos ^2(\theta ) \log ^2({ty^1})+ \csc ^2(\theta )}+\frac{t}{{y^1}}-\Omega
\label{y1ForBianchiVI}
,\end{equation}
\begin{equation}
{\Omega}=
2
\left(
\gamma_1
+\frac{\sum_q k_q\gamma_q}{k_1}
\right)
=\mbox{const}
,\end{equation}
where the constants $\gamma_i$ and $k_i$ are determined by the initial or boundary conditions for the source radiation.
Then other 
light path spatial coordinates
$y^p(t)$ will be determined by the relations (\ref{eqLightYp}) through the solution for $y^1(t)$ from relation (\ref{y1ForBianchiVI}).

From the equations (\ref{EqForGammaP})-(\ref{EqConnection}) for a gravitational wave in the Bianchi type VI universe follows a relation connecting the moment of radiation ${t'}$ and the variables defining the time $t$ and position of the observer detecting the electromagnetic field in gravitational wave: $z^1(t)=ty^1(t)$, $z^p(t)=ty^p(t)$.
The result relationship for ''the retarded time'' takes the following form in a synchronous reference frame:
\begin{equation}
t^2=
F\left(t',z^k(t)\right)
=
\frac{F_N\left(t',z^k(t)\right)}{F_D\left(t',z^1(t)\right)}
,\qquad
z^k(t)=ty^k(t)
\label{BianchiVItime1}
,\end{equation}
$$
F_N
=
\frac{
{y_{(s)}^1} {z^3}-{y_{(s)}^3} {z^1}
}{\sin (\theta ) 
}\,
({y_{(s)}^1} {t'}
{z^1}
)^{\sin (\theta )}
\Bigl[
({y_{(s)}^3} {z^1} - {y_{(s)}^1} {z^3})
\left(({y_{(s)}^1} {t'})^{\sin (\theta )}-\left({z^1}\right)^{\sin (\theta )}\right)
$$
$$
\mbox{}
+
\sin ^3(\theta ) \cos ^3(\theta ) \log ({z^1}) \log ({y_{(s)}^1} {t'}) 
({y_{(s)}^2} {z^1}-{y_{(s)}^1} {z^2}) (\log ({y_{(s)}^1} {t'})-\log ({z^1}))
$$
$$
\mbox{}
-\sin (\theta ) \cos (\theta ) ({y_{(s)}^2} {z^1}-{y_{(s)}^1} {z^2}) (\log ({y_{(s)}^1} {t'})-\log ({z^1}))
$$
$$
\mbox{}
-\sin ^2(\theta ) \cos ^2(\theta ) ({y_{(s)}^3} {z^1}
-{y_{(s)}^1} {z^3}) 
\left(\log ^2({y_{(s)}^1} {t'}) \left({z^1}\right)^{\sin (\theta )}-\log ^2({z^1}) ({y_{(s)}^1} {t'})^{\sin (\theta )}\right)
\Bigr]
$$
$$
\mbox{}
-\sin (\theta ) ({y_{(s)}^2} {z^1}-{y_{(s)}^1} {z^2}) 
\Bigl[
({y_{(s)}^2} {z^1}-{y_{(s)}^1} {z^2}) ({y_{(s)}^1} {t'})^{\sin (\theta )} \left(\cos ^2(\theta ) \log
   ^2({y_{(s)}^1} {t'})+\csc ^2(\theta )\right)
$$
$$
\mbox{}
-\left({z^1}\right)^{\sin (\theta )} 
\Bigl(
({y_{(s)}^2} {z^1}-{y_{(s)}^1} {z^2}) \left(\cos ^2(\theta ) \log
   ^2({z^1})+\csc ^2(\theta )\right)
$$
$$
\mbox{}
+({y_{(s)}^3} {z^1}-{y_{(s)}^1} {z^3}) ({y_{(s)}^1} {t'})^{\sin (\theta )} (\log ({y_{(s)}^1} {t'})-\log
   ({z^1})) 
\Bigl(
\cot (\theta )
$$
$$
\mbox{}
-\sin (\theta ) \cos ^3(\theta ) \log ({z^1}) \log ({y_{(s)}^1} {t'})
\Bigr)
\Bigr)
\Bigr]
$$
$$
\mbox{}
+{y_{(s)}^1} {t'} {z^1}^2
\Bigl[
\left(
({y_{(s)}^1} {t'})^{\sin (\theta )}-\left({z^1}\right)^{\sin (\theta )}
\right)^2
$$
\begin{equation}
\mbox{}
-\sin ^2(\theta ) \cos ^2(\theta ) 
({y_{(s)}^1}{t'}{z^1})^{\sin (\theta )} 
(\log ({y_{(s)}^1} {t'})-\log ({z^1}))^2
\Bigr]
\label{BianchiVItime2}
,\end{equation}
$$
F_D
=
{y_{(s)}^1}^2 {z^1} 
\Bigl[
\left(
({y_{(s)}^1} {t'})^{\sin (\theta)}-\left({z^1}\right)^{\sin (\theta )}
\right)^2
$$
\begin{equation}
\mbox{}
-\sin ^2(\theta ) \cos ^2(\theta )
 ({y_{(s)}^1} {t'}{z^1})^{\sin (\theta )} \left(\log ({y_{(s)}^1} {t'})-\log ({z^1})\right)^2
\Bigr]
\label{BianchiVItime3}
,\end{equation}
where $y^k_{(s)}=\left\{y_{(s)}^1,y_{(s)}^2,y_{(s)}^3\right\}$ specify the spatial position of the radiation source in a synchronous reference frame, $t'$ is the moment of radiation of the source, $t$ is the moment of detection of electromagnetic radiation by an observer having spatial coordinates $y^1(t)$, $y^2(t)$ and $y^3(t)$. The parameter $\theta$ is the angular parameter of the exact model of gravitational waves in the Bianchi type VI universe.


Explicit form of the gravitational wave metric in a synchronous reference frame and
relationships
(\ref{BianchiVItime1}), (\ref{BianchiVItime2}) and (\ref{BianchiVItime3}) which define 
the "retarded time''  of radiation in a gravitational wave in the Bianchi Type VI universe can be used to construct the retarded electromagnetic potentials of a charge in a gravitational wave and for other astrophysical problems related to the propagation of radiation in a gravitational wave.

\section{Conclusion}

For the space-time of a strong gravitational wave, explicit transformations from a privileged coordinate system to a synchronous reference system are found on the basis of the Hamilton-Jacobi formalism. The general form of the metric of a strong gravitational wave in a synchronous reference frame with separated time and space variables is obtained. In a synchronous reference frame, an explicit form of light trajectories in a gravitational wave is found. General relations have been constructed that determine the type of light ''cone'' of radiation propagation in a gravitational wave. The spatial components of the velocity of propagation of radiation in a gravitational wave in a synchronous reference frame are found. Relationships have been obtained for ''the retarded time'' during the propagation of electromagnetic radiation in a gravitational wave. A general form of relationships has been found connecting the time intervals of radiation from a source and the time intervals of detection of radiation by an observer in a gravitational wave.
As a particular example, relationships for the "retarded time" were obtained for the exact gravitational wave model in the Bianchi type VI universe.

\section{Acknowledgments}
The study was supported by the Russian Science Foundation, 
grant \mbox{No.\,23-22-00343}, 
 {https://rscf.ru/en/project/23-22-00343/}.



\end{document}